\documentclass[conference]{IEEEtran}
\IEEEoverridecommandlockouts
\usepackage{cite}
\usepackage{amsmath,amssymb,amsfonts}
\usepackage{algorithmic}
\usepackage{graphicx}
\usepackage{textcomp}
\usepackage{subfig}
\usepackage{xcolor}
\usepackage{booktabs}
\usepackage{siunitx} 
\DeclareUnicodeCharacter{2212}{-}

\def\BibTeX{{\rm B\kern-.05em{\sc i\kern-.025em b}\kern-.08em
    T\kern-.1667em\lower.7ex\hbox{E}\kern-.125emX}}
\begin{document}

\title{Exploring Lead Free Mixed Halide 
Double Perovskites Solar Cell\\}

\author{Md Yekra Rahman*, Dr. Sharif Mohammad Mominuzzaman\\ Department of Electrical \& Electronic Engineering\\Bangladesh University of Engineering and Technology, Dhaka, Bangladesh\\ *Corresponding Author Email: yekrarahman@gmail.com }
\maketitle

\begin{abstract}
The significant surge in energy use and escalating environmental concerns have sparked worldwide interest towards the study and implementation of solar cell technology. Perovskite solar cells (PSCs) have garnered remarkable attention as an emerging third-generation solar cell technology. This paper presents an in-depth analysis of lead-free mixed halide double perovskites in the context of their potential uses in solar cell technology. Through the previous studies of various mixed halide double perovskite materials as potential absorber layer materials, it has been observed that Cs\textsubscript{2}TiI\textsubscript{6−x}Br\textsubscript{x} possesses promising characteristics. In this study, simulations were conducted using SCAPS-1D software to explore all possible combinations for x= 0 to 6. The materials for the Hole Transport Layer (HTL) and Electron Transport Layer (ETL) and absorber layer, along with their optimal thicknesses, are selected to yield the most promising results in terms of open-circuit voltage (V$_{oc}$), short-circuit current density (J$_{sc}$), fill factor (FF), and power conversion efficiency. A novel tolerance factor is employed to assess structural stability of perovskites. FTO/Cu\textsubscript{2}O/Cs\textsubscript{2}TiI\textsubscript{5}Br\textsubscript{1}/WS\textsubscript{2} emerges as the best combination in terms of the efficiency with 19.03\% but FTO/Cu\textsubscript{2}O/Cs\textsubscript{2}TiI\textsubscript{2}Br\textsubscript{4}/WS\textsubscript{2} shows good stability with 13.31\% efficiency.\\

\end{abstract}

\section{Introduction}
Over the decade, solar cells made from materials like CdTe, CIGS, GaInP, and CZTS were outperformed by a new kind of solar cell called Perovskite Solar Cells (PSCs) \cite{green2021solar}. These PSCs are made from inorganic-organic halide ABX\textsubscript{3} type structure. It also possess several impressive and distinct photoelectric properties, such as a high absorption coefficient, long carrier diffusion length, high carrier mobility, low recombination rate, and tunable band gaps through chemical modification. \cite{Shi2015}. In the past few years, the efficiency in hybrid halide organic-inorganic perovskite solar cells, specifically CH\textsubscript{3}NH\textsubscript{3}PbI\textsubscript{3} (MAPbI\textsubscript{3}), has improved quickly, rising from 3.8\% in 2009 \cite{kojima2009organometal} to 26.1\% in 2023\cite{NREL}.

Despite the rapid increase in the efficiency, there are still a few major challenges hindering the widespread adoption of this technology. One issue is the intrinsic and extrinsic instability of the organic components, and another significant concern is the toxicity of lead. To address these issues, all-inorganic cesium halide perovskites are being explored as alternatives\cite{Ju2018}. In these, cesium Cs\textsuperscript{+} is used to replace MA\textsuperscript{+} or FA\textsuperscript{+} in the A-site of the perovskite structure. These are more stable thermodynamically and less prone to moisture absorption. Research by Yang et al. indicates that the high-temperature phase of CsPbBr\textsubscript{3} shows greater thermodynamic stability, as well as intrinsic and extrinsic stability\cite{Zhang2019}. To make perovskites more stable and lead-free on the B site, A\textsubscript{2}B\textsuperscript{I}B\textsuperscript{III}X\textsubscript{6} structures have been introduced. Building on this strategy, researchers have successfully synthesized several halide double perovskites, such as Cs\textsubscript{2}AgInCl\textsubscript{6} \cite{Volonakis2017} and Cs\textsubscript{2}AgSbCl\textsubscript{6} \cite{Tran2017}. But these new perovskites have excessively large band gaps. This makes them unsuitable for use in single-junction solar cells. Meanwhile, Woodward et. al have developed the first halide double perovskite Cs\textsubscript{2}AgBiBr\textsubscript{6} which has large band gap and resistance to air and water exposure \cite{Greenough2017}. After that, the mutation of B\textsuperscript{2+} ions in double perovskites has been achieved by incorporationg tetravalent cations resulting in "vacancy-ordered" double perovskites with the general formula A\textsubscript{2}BX\textsubscript{6}. Chen et al. successfully synthesized Cs\textsubscript{2}TiBr\textsubscript{6}, with experimental results showing that it possesses a stable bandgap of 1.8eV\cite{Ju2018}. The thin-film perovskite solar cells based on Cs\textsubscript{2}TiBr\textsubscript{6} have demonstrated a consistent efficiency of up to 3.3\%\cite{Chen2018}. Qui et al. conducted a detailed analysis of the band gaps, spectral characteristics, and formation energies of Cs\textsubscript{2}Ti(Br\textsubscript{1−x}Y\textsubscript{x})\textsubscript{6} (where Y represents Cl and I; and x varies from 0 to 1 in increments of 0.25) by utilizing first-principles methods based on Density Functional Theory (DFT) calculations \cite{li2020structure}.

Despite these developments, the challenge of identifying the most effective ionic substitutions for enhanced efficiency and stability remains. To forecast the stability of the perovskite structure based on the chemical formula ABX\textsubscript{3} Goldschmidt tolerance factor (t) has been used.

\begin{equation}
t=\frac{r_A+r_X}{\sqrt{2}(r_B+r_X)} \label{eq}
\end{equation} where ionic radii, \( r_i \), of each ion (i = A, B, X). For stable perovskite, \( 0.825 < t < 1.059 \) But It has only 74\% accuracy for 576 experimentally collected ABX\textsubscript{3} type materials\cite{Bartel2018}. To improve accuracy, a new tolerance factor is introduced by Bartel et al.
\begin{equation}
    \tau=\frac{r_X}{r_B}-n_A(n_A-\frac{\frac{r_A}{r_B}}{ln(\frac{r_A}{r_B})})
\end{equation} where \( n_A \) is the oxidation state. For stable perovskite, \( \tau < 4.18 \) and the accuracy is 92\% for the same database \cite{Bartel2018}.

In this study, we will further explore their effect in performance parameters of solar cell with Cs\textsubscript{2}TiI\textsubscript{6−x}Br\textsubscript{x} used as absorber layer. It also focuses on selecting specific materials for each of the other two key layers of PSCs, with the primary model being the FTO/ETL/Cs\textsubscript{2}TiI\textsubscript{2}Br\textsubscript{4}/HTL. Here Cs\textsubscript{2}TiI\textsubscript{2}Br\textsubscript{4} is chosen as initial absorber layer. Various permutations for this study are depicted in Fig-1, and modifications to the absorber perovskite layer are explored to assess their impact on performance metrics by using SCAPS-1D. Finally, the improved tolerance factor (\( \tau \)) has been used to the investigation of the absorber layer’s stability to assess the mixed halide double perovskite structure’s performance with various compositions.
\begin{figure}[htbp]
\centering
\includegraphics[width=.5\linewidth]{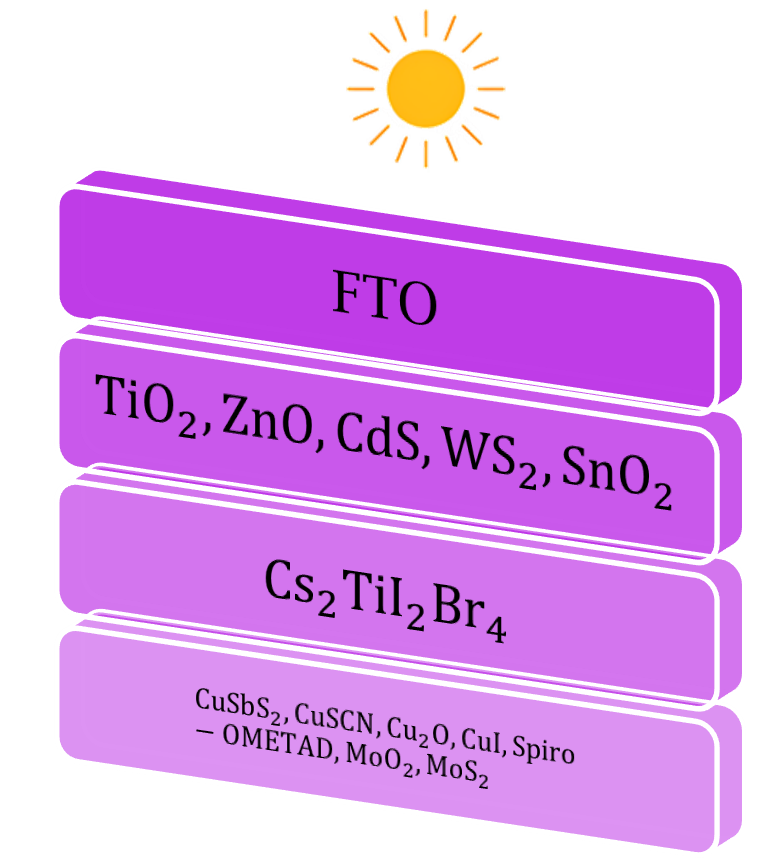}
\caption{Possible setups of cells }
\label{fig:enter-label}
\end{figure}

\section{Methodology} SCAPS-1D software requires all relevant factors for equation solving to be precisely configured. Key material parameters for the simulation include the bandgap (\(E_g\)) of each layer, electron affinity (\(\chi\)), relative dielectric permittivity (\(\epsilon_r\)), effective density of states in the conduction and valence bands (\(N_c\) and \(N_v\)), mobilities of electrons and holes (\(\mu_n\) and \(\mu_h\)), thermal velocities of electrons and holes (\(V_{th,n}\) and \(V_{th,h}\)). The calculation of the absorption coefficients for each layer was performed using the built-in model. For the front and back contacts of the solar cell, gold (Au) and aluminum (Al) were used, with their work functions set at 5.45 eV and 4.1 eV respectively. The defect density in each layer was assumed to be around \(10^{17} \text{ cm}^{-3}\). The cross-section for electron capture is measured at \(10^{-19} \text{ cm}^{-2}\) for Perovskite/ETL and \(10^{-18} \text{ cm}^{-2}\) for HTL/Perovskite. In contrast, the cross-section for hole capture stands at \(10^{-18} \text{ cm}^{-2}\) for Perovskite/ETL and \(10^{-19} \text{ cm}^{-2}\) for HTL/Perovskite respectively. Table I summarizes these parameters. The simulations were conducted under standard operating conditions using solar spectrum AM 1.5G and a temperature of 300K.

\begin{table*}[!t]
\caption{Initial material parameters used for SCAPS simulation \cite{islam2020simulation,matebese2018progress,jannat2021performance,hima2020enhancement,kohnehpoushi2018mos2}}
\label{your-table-label}
\centering
\scriptsize
\begin{tabular}{|l|l|l|l|l|l|l|l|l|l|l|}
\hline
Parameter & FTO & TiO$_2$ & Cs$_2$TiI$_2$Br$_4$ & CuSbS$_2$ & CuSCN & Cu$_2$O &  MoS$_2$ & CuI & WS$_2$&  ZnO  \\ \hline
Thickness ($\mu$m) & 0.5 & 0.05 & 0.35 & 0.35 & 0.35 & 0.35 &  0.35 & 0.35 & 0.05 & 0.05 \\ \hline
Bandgap (eV) & 3.5 & 3.2 & 1.38 & 1.58 & 3.68 & 2.17 &  1.29 & 2.98 & 1.87 & 3.3 \\ \hline
Electron affinity (eV) & 4.0 & 3.9  & 4.15 & 4.2 & 1.9 & 3 &  4.2 & 2.1 & 4.3 & 4.1 
\\ \hline
Dielectric permittivity & 9 & 38 & 6.6 & 8.2 & 10 & 7.5 &  3 & 6.5 & 11.9 &  9\\ \hline
Density of states at CB (cm$^{-3}$) & 2.2$\times10^{18}$ & 2.2$\times10^{18}$ & 6$\times10^{19}$ & 2$\times10^{18}$ & 2.2$\times10^{18}$ &
2$\times10^{18}$& 
2.2$\times10^{18}$& 
2.8$\times10^{19}$& 1$\times10^{19}$ &
2.2$\times10^{18}$ \\ \hline
Density of states at VB (cm$^{-3}$) & 1.8$\times10^{19}$ & 1.8$\times10^{19}$ & 2.4$\times10^{19}$ & 
$10^{19}$ &
1.8$\times10^{19}$ & 1.1$\times10^{19}$ &1.8$\times10^{19}$ & 1.1$\times10^{19}$ & 2.4$\times10^{19}$ & 1.9$\times10^{19}$ \\ \hline
Electron and Hole thermal velocity& 10$^{7}$ &
10$^{7}$ &
10$^{7}$ &
10$^{7}$ &
10$^{7}$ &
10$^{7}$ &
10$^{7}$ &
10$^{7}$ &
10$^{7}$ &
10$^{7}$ 
\\ \hline
Electron mobility (cm$^2$V$^{-1}$s$^{-1}$)& 20 & 
 20 & 
 8.16 & 
 4.9 & 
 10$^{-4}$ & 
 200 & 
 100 & 
 1.7$\times10^{-4}$ & 
 260 & 
 100 
\\ \hline
Hole mobility (cm$^2$V$^{-1}$s$^{-1}$)&10 & 
10 & 
8 & 
4.9 & 
10$^{-2}$& 
80 & 
 150 & 
 2$\times10^{-4}$ & 
 51 & 
 25  
\\ \hline
Donor Doping Density (cm$^{-3}$) & 
10$^{19}$ & 
10$^{16}$  &
10$^{18}$  &
0  &
0  &
0  &
0  &
0  &
10$^{17}$  &
10$^{17}$  
\\ \hline
Acceptor Doping Density (cm$^{-3}$) & 
0 & 
0 & 
10$^{18}$  &
10$^{18}$  &
10$^{17}$  &
2$\times10^{19}$ & 
10$^{18}$  &
10$^{18}$  &
0 & 
0 
\\ \hline
Defect Density (cm$^{-3}$) & 
10$^{15}$ &
10$^{17}$ &
10$^{15}$ &
10$^{17}$ &
10$^{17}$ &
10$^{17}$ &
10$^{17}$ &
10$^{17}$ &
10$^{17}$ &
10$^{17}$ 
\\ \hline
\end{tabular}
\end{table*}

\section{Results and Discussion}

\subsection{J-V Characteristics for different HTLs and varying their thickness}\label{AA}
The J-V characteristics for seven HTLs - \(\text{CuSbS}_2\), \(\text{CuSCN}\), \(\text{Cu}_2\text{O}\), \(\text{CuI}\), Spiro-OMETAD, \(\text{MoO}_2\), and \(\text{MoS}_2\) were investigated to assess their impact on cell performance. \(\text{CuSbS}_2\) shows the poorest performance, with minimum J$_{sc}$ and V$_{oc}$. So, it's not optimal for high-efficiency PSCs. \(\text{CuSCN}\), \(\text{Cu}_2\text{O}\), \(\text{CuI}\), Spiro-OMETAD, and \(\text{MoO}_2\) have similar J$_{sc}$ as displayed in Fig-2. However, \(\text{MoS}_2\) outperforms all in terms of J$_{sc}$. Spiro-OMETAD and \(\text{MoS}_2\) have also similar V$_{oc}$. But \(\text{Cu}_2\text{O}\) displays higher J$_{sc}$ and V$_{oc}$ for efficient charge transport and separation in PSCs. 
\begin{figure}[htbp]
    \centering
    \includegraphics[width=1\linewidth]{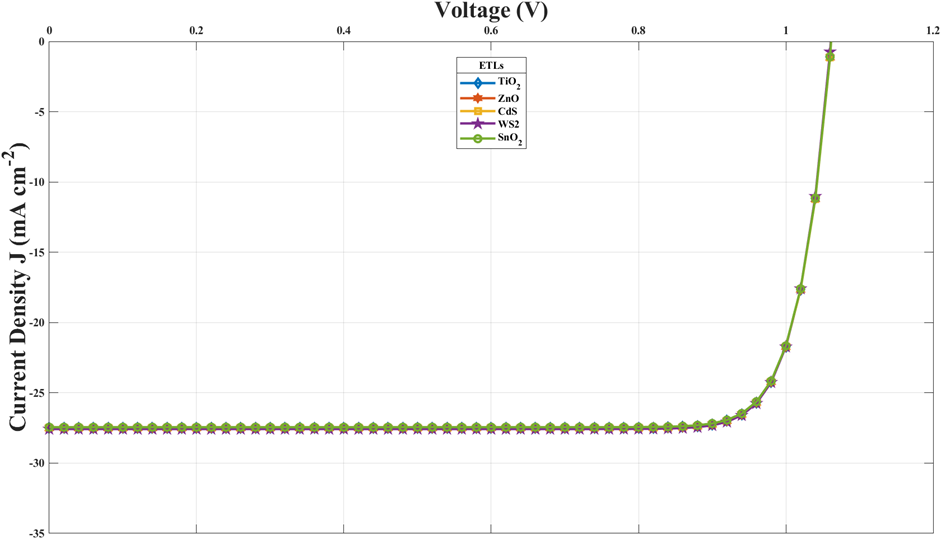}
    \caption{J-V characteristics for different HTLs}
    \label{fig:enter-label}
\end{figure}
Moreover, in terms of all thicknesses from 0.1$\mu m$ to 1$\mu m$ it is found that \(\text{Cu}_2\text{O}\) and \(\text{MoO}_2\) demonstrate higher efficiencies compared to other HTL materials (Fig-3a). \(\text{Cu}_2\text{O}\) maintains the highest efficiency and CuI shows a decline in efficiency with increasing thickness.
\begin{figure}[htbp]
    \centering
    \subfloat[]{\includegraphics[width=0.49\linewidth]{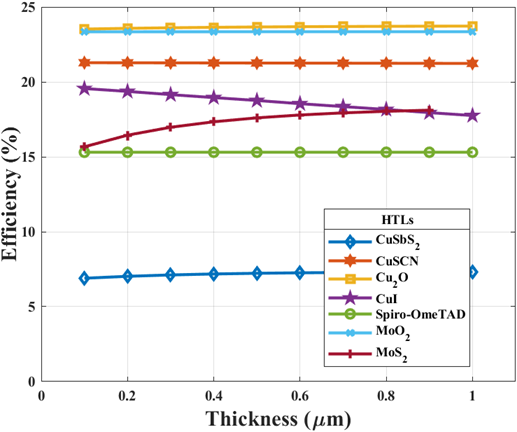}\label{fig:efficiency}}
    \hfil
    \subfloat[]{\includegraphics[width=0.49\linewidth]{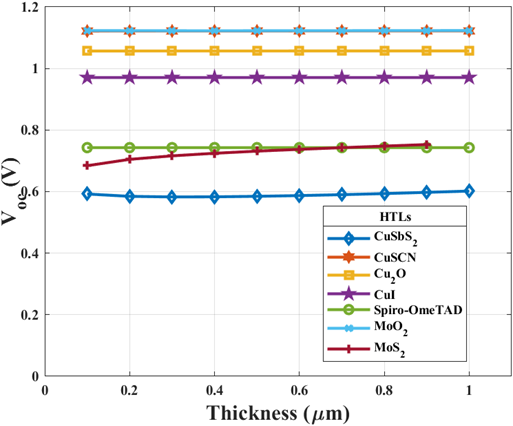}}\label{fig:voc}
    \hfil
    \subfloat[]{\includegraphics[width=0.49\linewidth]{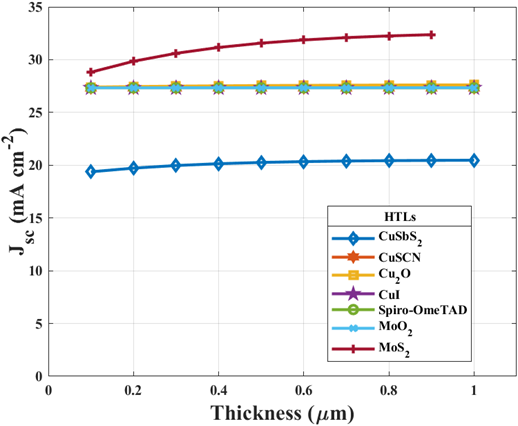}}\label{fig:jsc}
    \hfil
    \subfloat[]{\includegraphics[width=0.49\linewidth]{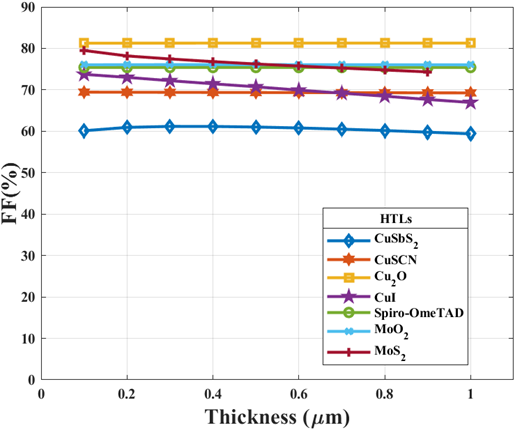}}\label{fig:ff}
    \caption{Performance parameters of various HTLs at different thicknesses. (a) Efficiency (b) Open-circuit voltage (c) Short-circuit current density (d) Fill factor}
    \label{fig:efficiency_fillfactor}
\end{figure}
\(\text{CuSbS}_2\) displays the lowest efficiency at all thicknesses.Furthermore, examining Fig-3(d), it can be observed 
that \(\text{Cu}_2\text{O}\)  exhibits the highest FF. Therefore, based on this assessment of efficiency and FF, \(\text{Cu}_2\text{O}\) was identified as the most appropriate HTL material for this study.

\subsection{J-V Characteristics for different ETLs and varying their thickness}\label{AA}
The J-V characteristics of five ETLs, specifically \(\text{TiO}_2\), \(\text{ZnO}\), \(\text{CdS}\), \(\text{WS}_2\), and \(\text{SnO}_2\), were evaluated to assess their efficacy in PSCs. By performing comparative analysis of the curves, as depicted in Fig-4, it can be seen that there is a lack of substantial variation among the ETLs. Consequently, determining an optimal ETL from the materials poses a challenge.
\begin{figure}[htbp]
    \centering
    \includegraphics[width=1\linewidth]{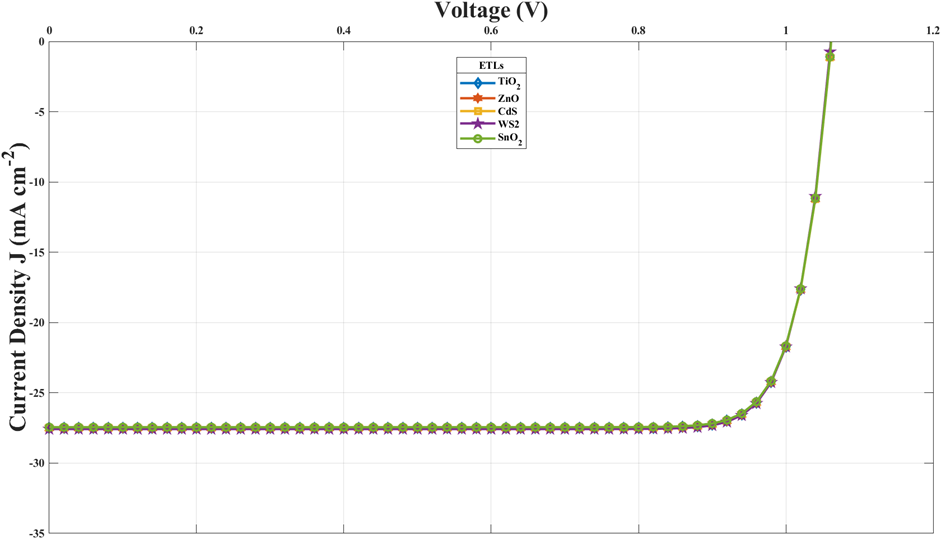}
    \caption{J-V characteristics for different ETLs}
    \label{fig:enter-label}
\end{figure}
The Y-axis of Fig-5 across all the curves indicates a negligible impact of layer thickness of the ETLs on the performance parameters. A closer look in the zoomed-in plot of Fig-5(a), reveals that \(\text{WS}_2\) shows higher efficiency at a thickness of 0.04 µm. Additionally,  \(\text{WS}_2\) also surpasses the other ETLs in terms of FF. These observations led to the selection of  \(\text{WS}_2\) as the preferred ETL material with the thickness of 0.04 µm. 


\begin{figure}[htbp]
    \centering
    \subfloat[]{\includegraphics[width=0.49\linewidth]{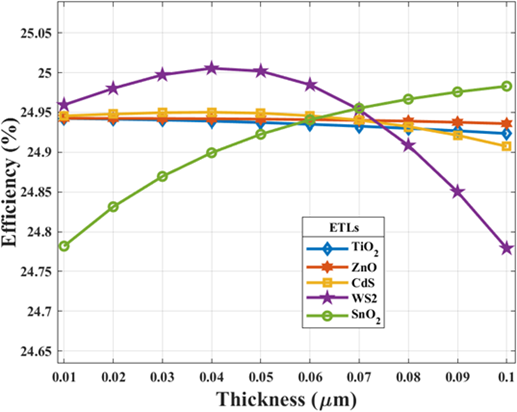}\label{fig:efficiency}}
    \hfil
    \subfloat[]{\includegraphics[width=0.49\linewidth]{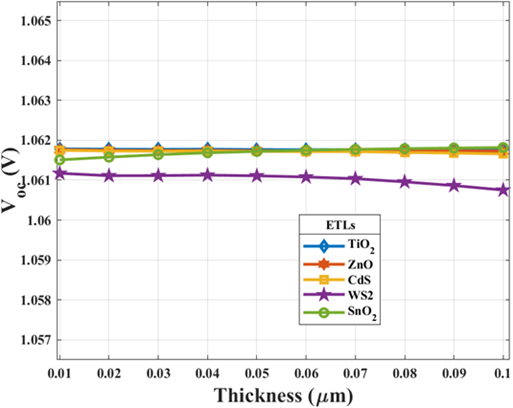}}\label{fig:voc}
    \hfil
    \subfloat[]{\includegraphics[width=0.49\linewidth]{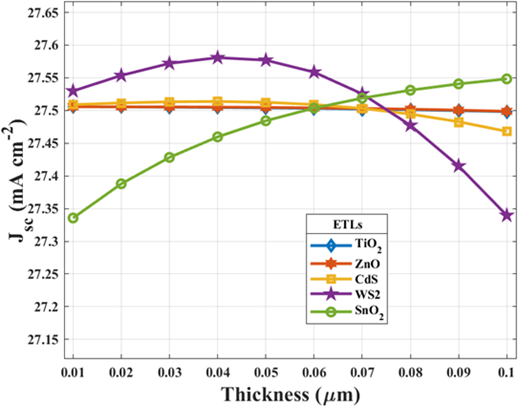}}\label{fig:jsc}
    \hfil
    \subfloat[]{\includegraphics[width=0.49\linewidth]{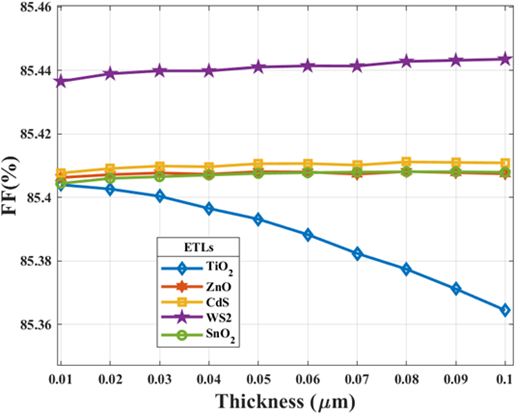}}\label{fig:ff}
    \caption{Performance parameters of various ETLs at different thicknesses. (a) Efficiency (b) Open-circuit voltage (c) Short-circuit current density (d) Fill factor}
    \label{fig:efficiency_fillfactor}
\end{figure}

\subsection{Effect of donor and acceptor concentration of ETL, HTL and absorber layer }
Cu\textsubscript{2}O and WS\textsubscript{2} were identified as the respective ETL, HTL from the previous observations. Cs$_2$TiI$_2$Br$_4$ was selected as initial absorber layer . Four 2D color maps in each of Fig-6 and Fig-7 delineate the variations in efficiency, V$_{oc}$, J$_{sc}$, and FF, with color gradations ranging from deep blue (low values) to yellow (high values). 
\begin{figure}[htbp]
    \centering
    \subfloat[]{\includegraphics[width=0.49\linewidth]{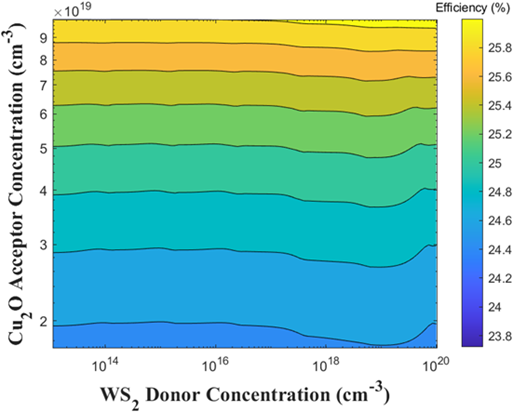}\label{fig:efficiency}}
    \hfil
    \subfloat[]{\includegraphics[width=0.49\linewidth]{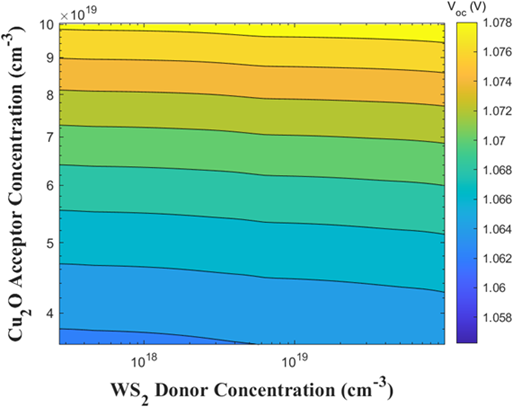}}\label{fig:voc}
    \hfil
    \subfloat[]{\includegraphics[width=0.49\linewidth]{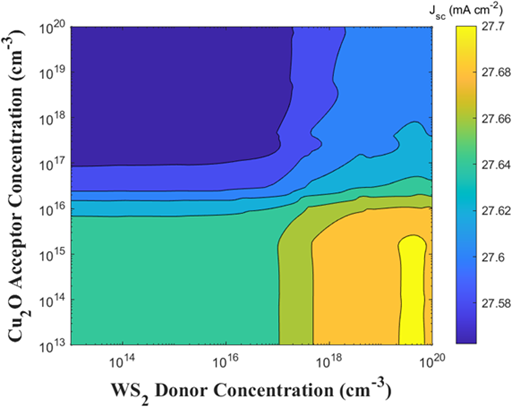}}\label{fig:jsc}
    \hfil
    \subfloat[]{\includegraphics[width=0.49\linewidth]{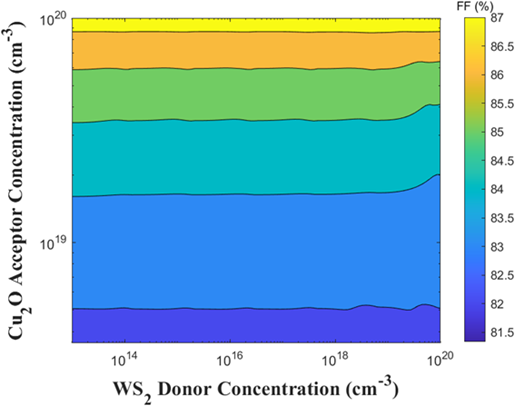}}\label{fig:ff}
    \caption{Performance parameters for varying acceptors and donors concentration of Cu\textsubscript{2}O and WS\textsubscript{2}. (a) Efficiency (b) Open-circuit voltage (c) Short-circuit current density (d) Fill factor}
    \label{fig:efficiency_fillfactor}
\end{figure}
Fig-6(a) illustrates that higher efficiency can be attained by increasing both the acceptor and donor concentration in Cu\textsubscript{2}O and WS\textsubscript{2} respectively. Fig-6(c) indicates that lower Cu\textsubscript{2}O acceptor concentrations and higher WS\textsubscript{2} donor concentrations show higher J$_{sc}$.  Moreover, as shown in Fig-6(d), FF is more influenced by Cu\textsubscript{2}O’s acceptor concentration. So, this study adopts an acceptor concentration of 8.5$\times10^{19}$ cm$^{-3}$ for Cu\textsubscript{2}O and a donor concentration of 5$\times10^{18}$ cm$^{-3}$ for WS\textsubscript{2}.
\begin{figure}[htbp]
    \centering
    \subfloat[]{\includegraphics[width=0.49\linewidth]{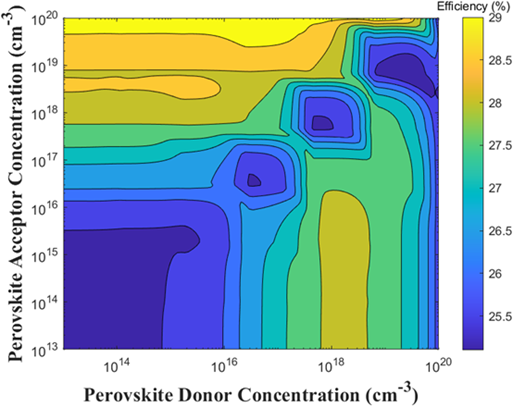}\label{fig:efficiency}}
    \hfil
    \subfloat[]{\includegraphics[width=0.49\linewidth]{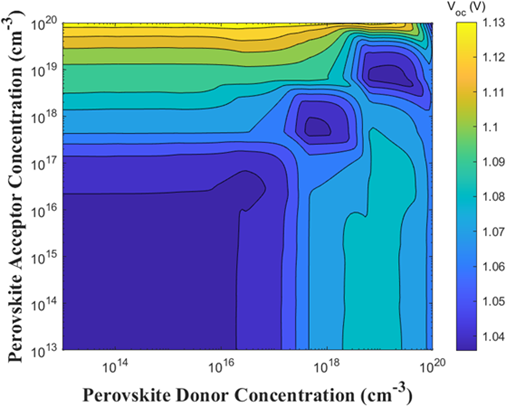}}\label{fig:voc}
    \hfil
    \subfloat[]{\includegraphics[width=0.49\linewidth]{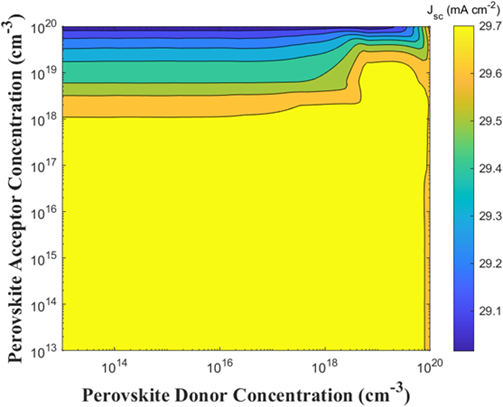}}\label{fig:jsc}
    \hfil
    \subfloat[]{\includegraphics[width=0.49\linewidth]{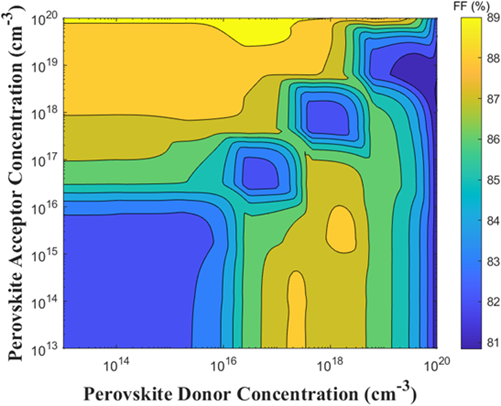}}\label{fig:ff}
    \caption{Performance parameters for varying acceptors and donors concentration of absorber layer. (a) Efficiency (b) Open-circuit voltage (c) Short-circuit current density (d) Fill factor}
    \label{fig:efficiency_fillfactor}
\end{figure}

As depicted in Fig-7(a), It can be noted that the higher efficiency for  Cs$_2$TiI$_2$Br$_4$ is achieved when the acceptor concentration is around 10$^{19}$ and the donor concentration is within the vicinity of 10$^{17}$. Fig-7(d) indicates that a high FF can be achieved with an acceptor concentration of about 10$^{19}$ cm$^{-3}$ and a donor concentration ranging between 10$^{16}$ cm$^{-3}$ and 10$^{17}$  cm$^{-3}$. From the insights gained from these analyses, the optimal concentration for the absorber layer was set with an acceptor concentration of 8.9$\times10^{19}$ cm$^{-3}$ and a donor concentration of 4.5$\times10^{17}$ cm$^{-3}$.

\subsection{Effect of thickness and defect density of absorber layer }
Fig-8 and Fig-9 show how the thickness and defect density of the absorber layer affects solar cell performance. 
\begin{figure}[htbp]
    \centering
    \subfloat[]{\includegraphics[width=0.49\linewidth]{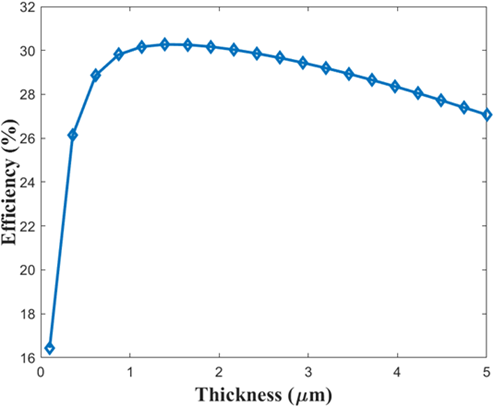}\label{fig:efficiency}}
    \hfil
    \subfloat[]{\includegraphics[width=0.49\linewidth]{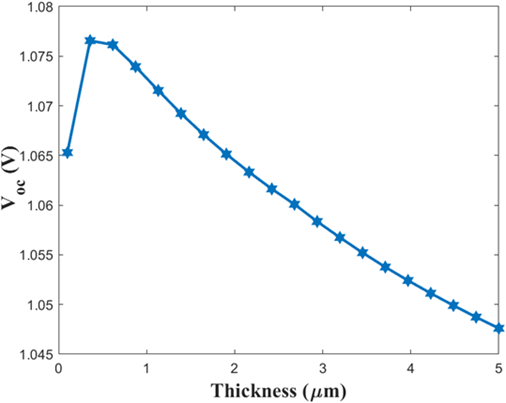}}\label{fig:voc}
    \hfil
    \subfloat[]{\includegraphics[width=0.49\linewidth]{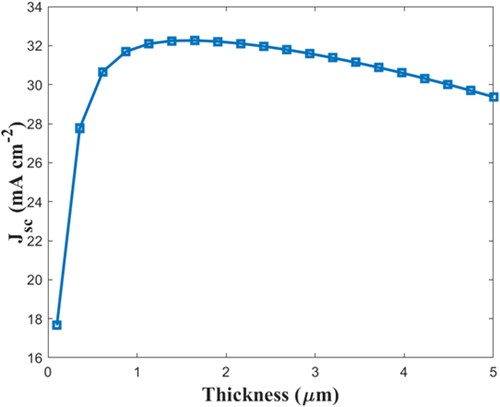}}\label{fig:jsc}
    \hfil
    \subfloat[]{\includegraphics[width=0.49\linewidth]{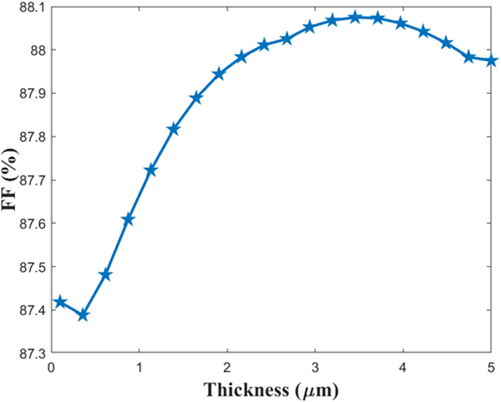}}\label{fig:ff}
    \caption{Performance parameters for varying thickness of absorber layer. (a) Efficiency (b) Open-circuit voltage (c) Short-circuit current density (d) Fill factor}
    \label{fig:efficiency_fillfactor}
\end{figure}

\begin{figure}[htbp]
    \centering
    \subfloat[]{\includegraphics[width=0.49\linewidth]{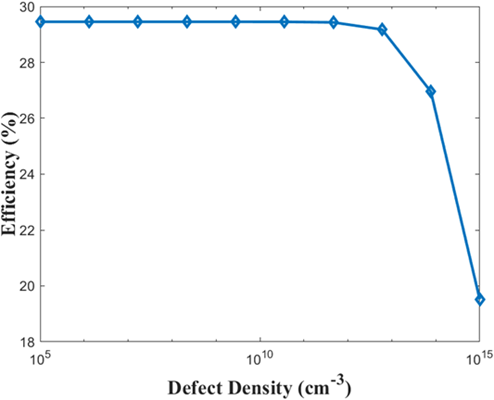}\label{fig:efficiency}}
    \hfil
    \subfloat[]{\includegraphics[width=0.49\linewidth]{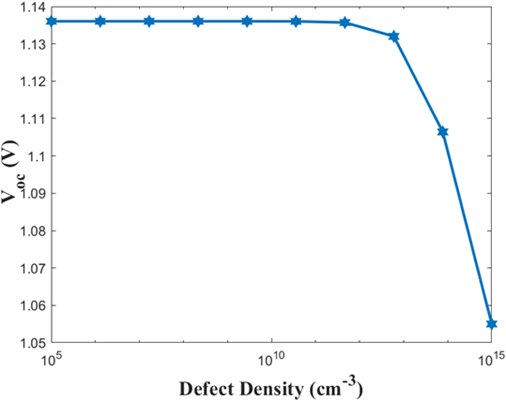}}\label{fig:voc}
    \hfil
    \subfloat[]{\includegraphics[width=0.49\linewidth]{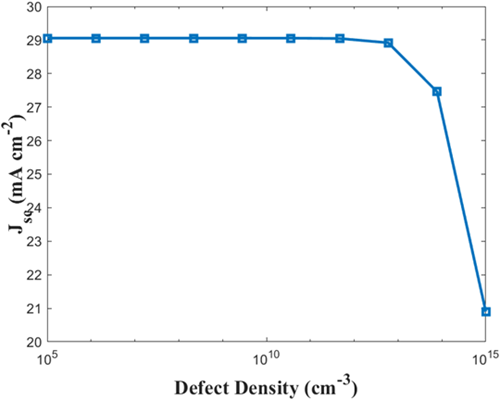}}\label{fig:jsc}
    \hfil
    \subfloat[]{\includegraphics[width=0.49\linewidth]{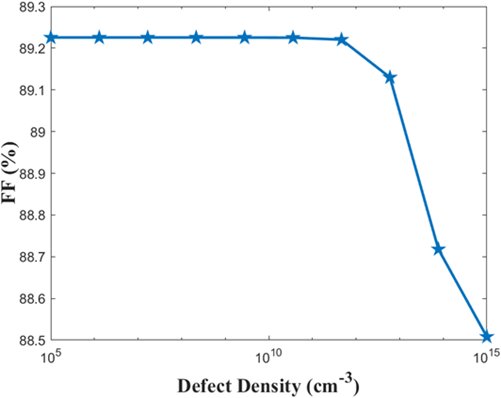}}\label{fig:ff}
    \caption{Performance parameters for varying defect density of absorber layer. (a) Efficiency (b) Open-circuit voltage (c) Short-circuit current density (d) Fill factor}
    \label{fig:efficiency_fillfactor}
\end{figure}

In Fig-8(a), an increase in efficiency acn be seen with the increase of absorber layer thickness up to a certain point. After that the efficiency begins to diminish. Likewise, an increase in FF up to a thickness of about 3.2 µm, followed by a subsequent decrease can be seen from Fig-8(d). On the other hand, the efficiency remains relatively unaffected until a defect density of 10$^{12}$ cm$^{-3}$ is reached. Beyond that a significant drop in efficiency occurs (Fig-9(a)). It notably falls from around 29\% to 19\% as the defect density escalates from 10$^{12}$ cm$^{-3}$  to 10$^{15}$ cm$^{-3}$ . This highlights cell's susceptibility to higher defect densities in absorber layer. Forthis study, a defect density of 10$^{15}$ cm$^{-3}$ was selected based on practical considerations and empirical validation.

In contrast, the FF's response to increasing defect densities, as illustrated in Fig-9(d), is less severe. The FF decreases marginally from 89.2\% to 88.5\% when defect density rises from 10$^{12}$  to 10$^{15}$ , indicating a relatively resilient behavior of FF against variations in defect density within the specified range. Although the decrease in FF should not be overlooked, its impact is less significant compared to the effect on 
efficiency.

\subsection{J-V characteristics and spectral responsivity for Cs\textsubscript{2}TiI\textsubscript{6−x}Br\textsubscript{x} (x=0 to 6)}
The influence of varying the mixed halide composition (denoted as x) in Cs\textsubscript{2}TiI\textsubscript{6−x}Br\textsubscript{x}, ranging from x=0 to 6, on the performance of the solar cell has been showcased in Fig-10. The analysis of V$_{oc}$ reveals that altering x between 1 and 5 results in minimal V$_{oc}$ variation. However, at the extremes of x=0 and 6, there is a noticeable change in V$_{oc}$, indicating that these materials significantly influence V$_{oc}$. 

\begin{figure}[htbp]
    \centering
    \includegraphics[width=1\linewidth]{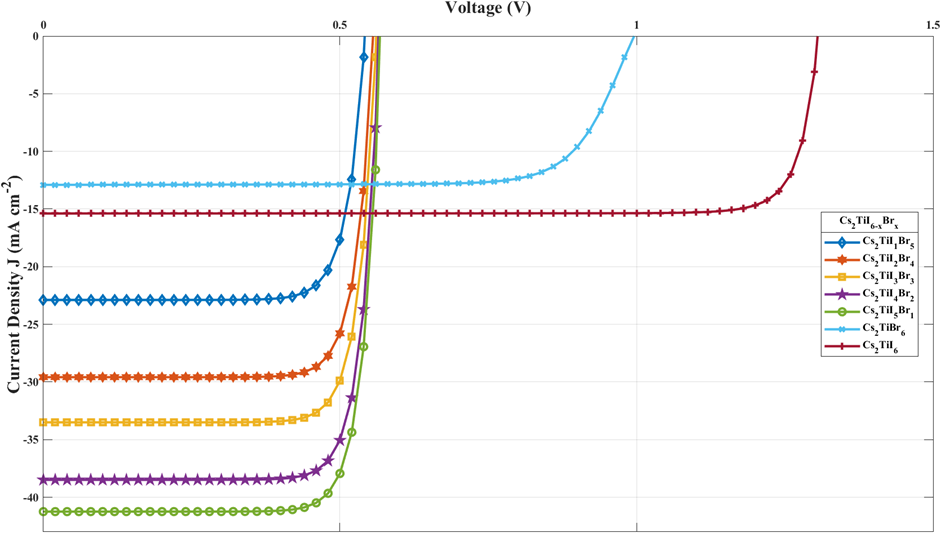}
    \caption{J-V characteristics for Cs\textsubscript{2}TiI\textsubscript{6−x}Br\textsubscript{x} (x= 0 to 6)}
    \label{fig:enter-label}
\end{figure}

In terms of J$_{sc}$, a trend of decreasing J$_{sc}$ with an increase in x from 1 to 5 was observed to generate electron hole pairs. The combination of these findings leads to the deduction that Cs\textsubscript{2}TiI\textsubscript{5}Br\textsubscript{1} possesses the most efficient absorber layer characteristics, evidenced by the stable V$_{oc}$ and the optimal J$_{sc}$

\begin{figure}[htbp]
    \centering
    \includegraphics[width=1\linewidth]{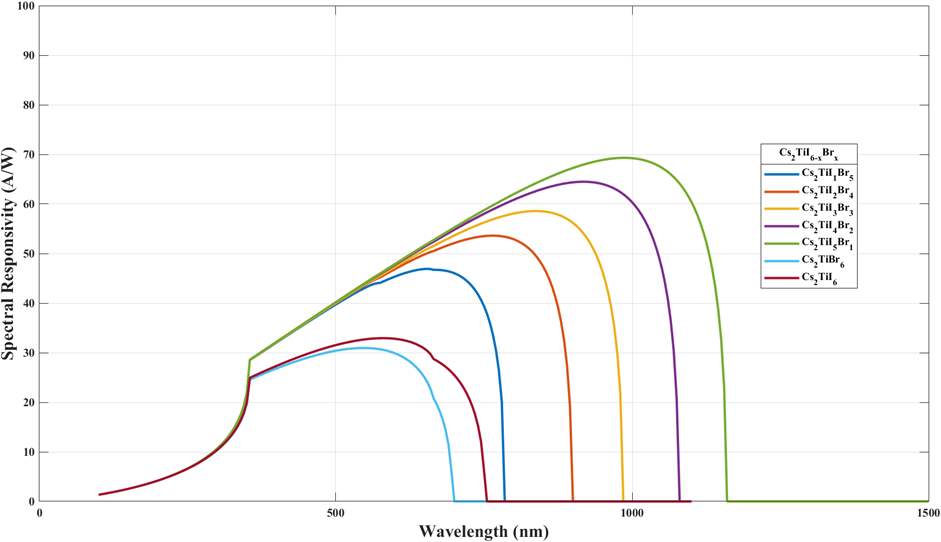}
    \caption{Spectral responsivity for Cs\textsubscript{2}TiI\textsubscript{6−x}Br\textsubscript{x} (x= 0 to 6)}
    \label{fig:enter-label}
\end{figure}
This study also explored the spectral responsivity. Fig-11 presents the spectral responsivity of Cs\textsubscript{2}TiI\textsubscript{6−x}Br\textsubscript{x} (x= 0 to 6). A notable observation is that as there is a rise in bromine concentration, there is a shift in the peak response wavelength towards the shorter side of the spectrum. Furthermore, with the decrease in x, the amplitude of this peak also increases.
\subsection{Performance and stability analysis}

This paper's focus on mixed halide double perovskite materials involved a comparative analysis of their efficiency and stability to identify the most suitable absorber material for PSCs. The key findings are summarized in Table II.

It can be seen that an increase in the value of x in (Cs\textsubscript{2}TiI\textsubscript{6−x}Br\textsubscript{x}) leads to a reduction in efficiency. The material Cs\textsubscript{2}TiI\textsubscript{5}Br\textsubscript{1} showcased the highest efficiency among the variants tested. However, its stability, when assessed using the improved tolerance factor ($\tau$) and the associated stability probability P($\tau$), is significantly lower compared to other materials.

On the other hand, materials like Cs\textsubscript{2}TiBr\textsubscript{6}, Cs\textsubscript{2}TiI\textsubscript{1}Br\textsubscript{5}, and Cs\textsubscript{2}TiI\textsubscript{2}Br\textsubscript{4}, which have lower efficiencies, exhibit higher stability. Notably, Cs\textsubscript{2}TiI\textsubscript{2}Br\textsubscript{4} achieves a balance between efficiency (13.31\%) and stability, displaying the highest efficiency among the more stable materials.

Therefore, Cs\textsubscript{2}TiI\textsubscript{2}Br\textsubscript{4} is identified as a superior absorber material considering both efficiency and stability and emerges as a promising candidate for future research and potential application in perovskite solar cell technology.

\begin{table*}[!t]
\centering
\caption{Performance Parameters with $\tau$ and P($\tau$) for Various Mixed Halide Double Perovskites}
\label{tab:perovskite_parameters}
\begin{tabular}{Slllllll}
\toprule
& \textbf{
Cs\textsubscript{2}TiBr\textsubscript{6}} & \textbf{Cs\textsubscript{2}TiI\textsubscript{1}Br\textsubscript{5}} & \textbf{Cs\textsubscript{2}TiI\textsubscript{2}Br\textsubscript{4}} & \textbf{Cs\textsubscript{2}TiI\textsubscript{3}Br\textsubscript{3}} & \textbf{Cs\textsubscript{2}TiI\textsubscript{4}Br\textsubscript{2}} & 
\textbf{Cs\textsubscript{2}TiI\textsubscript{5}Br\textsubscript{1}} & \textbf{Cs\textsubscript{2}TiI\textsubscript{6}} \\
\midrule
\textbf{V\textsubscript{oc} (V)} & 0.995876 & 0.542262 & 0.556572 & 0.56144 & 0.56598 & 0.568603 & 1.306315 \\
\textbf{J\textsubscript{sc} (mA/cm\textsuperscript{2})} & 12.90499 & 22.88736 & 29.58543 & 33.50023 & 38.48739& 41.24266 & 15.38558 \\
\textbf{FF (\%)}  & 77.4376  & 80.1015  & 80.864 & 81.0635 & 81.167 & 81.17 & 87.7209  \\
\textbf{Efficiency (\%)}& 9.9521   & 9.9414   & 13.3154 & 15.2467 & 17.6807&19.0349& 17.6305  \\
\textbf{$\tau$}& 4.074187 & 4.120698 & 4.167210 & 4.213722& 4.260233 & 4.306745 & 4.353257 \\
\textbf{P($\tau$)} & 0.584539 & 0.562502& 0.540138 & 0.517977 & 0.495071& 0.473635 & 0.450955 \\
\bottomrule
\end{tabular}
\end{table*}
\section{Conclusion}
In this study, we analyzed the performance and stability of materials used in ETLs, HTLs, and absorber layer of PSCs. Our investigation revealed CuSbS\textsubscript{2} as the least effective ETL, while Cu\textsubscript{2}O and MoO\textsubscript{2} outperformed other HTLs, with Cu\textsubscript{2}O displaying the highest efficiency across various thicknesses. Optimal charge carrier concentrations were determined, with 8.5$\times10^{19}$ cm$^{-3}$ for the acceptor in Cu\textsubscript{2}O and 5$\times10^{18}$ cm$^{-3}$ for the donor in WS\textsubscript{2}. A critical outcome was the impact of absorber layer thickness on device performance. Defect density also emerged as a significant factor, affecting efficiency more severely than FF. Chosen defect density of 10$^{15}$ cm$^{-3}$ in the absorber layer  balanced efficiency and stability considerations. Furthermore, we explored mixed halide double perovskites, pinpointing Cs\textsubscript{2}TiI\textsubscript{5}Br\textsubscript{1} as a potential efficient absorber with a wide absorption range up to 1155 nm. 
In the stability analysis Cs\textsubscript{2}TiI\textsubscript{2}Br\textsubscript{4} emerged as a promising candidate, offering a balance of high efficiency and good stability. This study provides insights for developing high-performance and stable PSCs which will contribute significantly to the advancement of effective solar energy solutions.

\bibliographystyle{ieeetr}
\bibliography{Yekra_Conference_01}

\begin{thebibliography}{10}

\bibitem{green2021solar}
M.~Green, E.~Dunlop, J.~Hohl-Ebinger, M.~Yoshita, N.~Kopidakis, and X.~Hao, ``Solar cell efficiency tables (version 57),'' {\em Progress in photovoltaics: research and applications}, vol.~29, no.~1, pp.~3--15, 2021.

\bibitem{Shi2015}
D.~Shi, V.~Adinolfi, R.~Comin, M.~Yuan, E.~Alarousu, A.~Buin, Y.~Chen, S.~Hoogland, A.~Rothenberger, K.~Katsiev, Y.~Losovyj, X.~Zhang, P.~A. Dowben, O.~F. Mohammed, E.~H. Sargent, and O.~M. Bakr, ``Low trap-state density and long carrier diffusion in organolead trihalide perovskite single crystals,'' {\em Science}, vol.~347, pp.~519--522, 1 2015.

\bibitem{kojima2009organometal}
A.~Kojima, K.~Teshima, Y.~Shirai, and T.~Miyasaka, ``Organometal halide perovskites as visible-light sensitizers for photovoltaic cells,'' {\em Journal of the american chemical society}, vol.~131, no.~17, pp.~6050--6051, 2009.

\bibitem{NREL}
NREL, ``Best research-cell efficiency chart,'' {\em U.S. Department of Energy, Office of Energy Efficiency and Renewable Energy}, 2024 https://www.nrel.gov/pv/interactive-cell-efficiency.html.

\bibitem{Ju2018}
M.~G. Ju, M.~Chen, Y.~Zhou, H.~F. Garces, J.~Dai, L.~Ma, N.~P. Padture, and X.~C. Zeng, ``Earth-abundant nontoxic titanium(iv)-based vacancy-ordered double perovskite halides with tunable 1.0 to 1.8 ev bandgaps for photovoltaic applications,'' {\em ACS Energy Letters}, vol.~3, pp.~297--304, 2 2018.

\bibitem{Zhang2019}
X.~Zhang, L.~Li, Z.~Sun, and J.~Luo, ``Rational chemical doping of metal halide perovskites,'' {\em Chemical Society Reviews}, vol.~48, pp.~517--539, 1 2019.

\bibitem{Volonakis2017}
G.~Volonakis, A.~A. Haghighirad, R.~L. Milot, W.~H. Sio, M.~R. Filip, B.~Wenger, M.~B. Johnston, L.~M. Herz, H.~J. Snaith, and F.~Giustino, ``Cs2inagcl6: A new lead-free halide double perovskite with direct band gap,'' {\em Journal of Physical Chemistry Letters}, vol.~8, pp.~772--778, 2 2017.

\bibitem{Tran2017}
T.~T. Tran, J.~R. Panella, J.~R. Chamorro, J.~R. Morey, and T.~M. McQueen, ``Designing indirect–direct bandgap transitions in double perovskites,'' {\em Materials Horizons}, vol.~4, pp.~688--693, 7 2017.

\bibitem{Greenough2017}
T.~Ibn-Mohammed, S.~C. Koh, I.~M. Reaney, A.~Acquaye, G.~Schileo, K.~B. Mustapha, and R.~Greenough, ``Perovskite solar cells: An integrated hybrid lifecycle assessment and review in comparison with other photovoltaic technologies,'' {\em Renewable and Sustainable Energy Reviews}, vol.~80, pp.~1321--1344, 12 2017.

\bibitem{Chen2018}
M.~Chen, M.~G. Ju, A.~D. Carl, Y.~Zong, R.~L. Grimm, J.~Gu, X.~C. Zeng, Y.~Zhou, and N.~P. Padture, ``Cesium titanium(iv) bromide thin films based stable lead-free perovskite solar cells,'' {\em Joule}, vol.~2, pp.~558--570, 3 2018.

\bibitem{li2020structure}
W.~Li, S.~Zhu, Y.~Zhao, and Y.~Qiu, ``Structure, electronic and optical properties of cs2ti (br1-xyx) 6 (y= cl, i; x= 0, 0.25, 0.5, 0.75, 1) perovskites: the first principles investigations,'' {\em Journal of Solid State Chemistry}, vol.~284, p.~121213, 2020.

\bibitem{Bartel2018}
C.~J. Bartel, C.~Sutton, B.~R. Goldsmith, R.~Ouyang, C.~B. Musgrave, L.~M. Ghiringhelli, and M.~Scheffler, ``New tolerance factor to predict the stability of perovskite oxides and halides,'' {\em Science Advances}, vol.~5, 1 2018.

\bibitem{islam2020simulation}
T.~Islam, R.~Jani, S.~M. Al~Amin, K.~M. Shorowordi, S.~S. Nishat, A.~Kabir, M.~Taufique, S.~Chowdhury, S.~Banerjee, and S.~Ahmed, ``Simulation studies to quantify the impacts of point defects: an investigation of cs2agbibr6 perovskite solar devices utilizing zno and cu2o as the charge transport layers,'' {\em Computational Materials Science}, vol.~184, p.~109865, 2020.

\bibitem{matebese2018progress}
F.~Matebese, R.~Taziwa, and D.~Mutukwa, ``Progress on the synthesis and application of cuscn inorganic hole transport material in perovskite solar cells,'' {\em Materials}, vol.~11, no.~12, p.~2592, 2018.

\bibitem{jannat2021performance}
F.~Jannat, S.~Ahmed, and M.~A. Alim, ``Performance analysis of cesium formamidinium lead mixed halide based perovskite solar cell with moox as hole transport material via scaps-1d,'' {\em Optik}, vol.~228, p.~166202, 2021.

\bibitem{hima2020enhancement}
A.~Hima and N.~Lakhdar, ``Enhancement of efficiency and stability of ch3nh3gei3 solar cells with cusbs2,'' {\em Optical Materials}, vol.~99, p.~109607, 2020.

\bibitem{kohnehpoushi2018mos2}
S.~Kohnehpoushi, P.~Nazari, B.~A. Nejand, and M.~Eskandari, ``Mos2: a two-dimensional hole-transporting material for high-efficiency, low-cost perovskite solar cells,'' {\em Nanotechnology}, vol.~29, no.~20, p.~205201, 2018.

\end{thebibliography}
\end{document}